\documentclass[slac_one]{revtex4}
\usepackage{graphicx}
\usepackage{fancyhdr}
\begin{document}

\title{CP violating anomalous tau lepton coupling $\bar{\tau}\tau V$ via a spin-0 unparticle
} 

%

\author{ \bf A. Moyotl}
\email[E-mail:]{amoyotl@ifuap.buap.mx}
\affiliation{\sl Instituto de F\'isica,  Benem\'erita Universidad
Aut\'onoma de Puebla, Apartado Postal 72570 Puebla, M\'exico}

\begin{abstract}
The tau  is the only known lepton that can decay hadronically. From this class of decays, high precision measurements of several quantities can be
extracted: the CKM matrix element $V_{us}$, the mass of the
strange quark, etc. Also, as a result of its large variety of decay channels,
the study of the tau lepton represents an interesting tool to search for CP
violation and other new physics effects. In this work, a spin-0 unparticle is considered and its contributions to the electromagnetic and weak properties
of the tau lepton are calculated. It is found that such contributions can be greater than the standard model (SM) contributions, though they can be of similar order of magnitude than the contributions from other SM extensions, such as an extension of the minimal supersymmetric standard model with vectorlike multiplets. For the unparticles parameters, the most recent CMS bound
from mono-jet production plus missing transverse energy at the LHC are considered.
\end{abstract}

\maketitle

\thispagestyle{fancy}


\section{Introduction} 

The origin of CP violation remains  an unsolved problem since the discovery of $CP$ violation in the $K^0-\bar{K}^0$ system \cite{Christenson:1979tu}. Although the observed CP violation in the $K$ meson system can be explained in the standard model (SM) due to a complex phase in the Cabibbo-Kobayashi-Maskawa (CKM) matrix \cite{Kobayashi:1973fv}, it is not clear if the CKM mechanism is correct and if it is the only source of CP or T violation. However, recent evidences of neutrino oscillations \cite{Fukuda:1998mi} suggest that these particles are massive, which opens up the possibility for lepton flavor violation and also for a source of CP violation in the lepton sector. A very promising scenario for large effects of CP violation arises from the electric (EDM) and weak electric dipole moments (WEDM) of the $\tau$ lepton. As far as the $\tau$ EDM is concerned, the current constraint was obtained by the  Belle collaboration at the KEKB collider using 29.5 fb$^{-1}$ of  data from the $e^+e^- \to \tau^+\tau^-$ process at $\sqrt{s}=10{.}58$ GeV \cite{Inami:2002ah}: the allowed limits are $-0{.}220<  \text{Re}(d_\tau)\times10^{16}/e\text{ cm} < 0{.}45$ and $-0{.}250 < \text{Im}(d_\tau)\times10^{16}/e\text{ cm} < 0{.}080$. On the other hand, the current bounds  of the $\tau$ WEDM were obtained by the  ALEPH collaboration \cite{Heister:2002ik} using data  from the process $e^+e^- \to \tau^+\tau^-$ collected from 1990 to 1995 at an energy of $\sqrt{s}=m_Z$ and an integrated luminosity of 155 pb$^{-1}$: $\text{Re}(d_\tau^{W}) < 0{.}5 \times 10^{-17} e\text{ cm}$ and $\text{Im}(d_\tau^{W})<  1{.}1 \times 10^{-17} e\text { cm}$. These results are much larger than the theoretical predictions of the SM, where $d_\tau^{\text{SM}}<10^{-34}$ e cm \cite{Hoogeveen:1990cb} and  $d_\tau^{W}<8\times 10^{-34}$ e cm \cite{Booth:1993af}. However, several SM extensions do predict large contributions to these properties, which in fact are closer to the experimental bounds. These properties will be explored below in the unparticle physics framework \cite{Georgi:2007ek}.

\section{Unparticles Physics overview}

Banks and Zaks analyzed a  model with a scale invariant sector in \cite{Banks:1981nn}, but it was only after the work of Georgi \cite{Georgi:2007ek} that high energy physicists  became more interested in this idea. Unparticle physics assumes the existence of a scale invariant hidden sector, known as
${\mathcal B}{\mathcal Z}$ sector, which can interact with the SM fields via the exchange of very heavy particles at a very high energy scale
${\mathcal M}_{\mathcal U}$. Below this energy scale, there are nonrenormalizable couplings between the fields of the ${\mathcal B}{\mathcal Z}$ sector and the SM ones. These couplings can be written generically as
$ {\mathcal O}_{SM}{\mathcal O}_{{\mathcal B}{\mathcal Z}}/{\mathcal M}_{\mathcal U}^{d_{SM}+d_{{\mathcal B}{\mathcal Z}}-4}$.  The dimension of the associated operators are $d_{{\mathcal B}{\mathcal Z}}$ and $d_{SM}$, respectively. Dimensional transmutation occurs at an energy scale $\Lambda_{\mathcal U}$ due to the renormalizable couplings of the ${\mathcal B}{\mathcal Z}$ sector. Below the scale $\Lambda_{\mathcal U}$, an effective theory can be used to describe the interactions between the fields of the ${\mathcal B}{\mathcal Z}$ sector and the SM fields, which arise from the exchange of unparticle fields. The effective Lagrangian describing the scalar and pseudo-scalar interactions of a spin-0 unparticle with a fermion pair is given by:

\begin{equation}\label{lup}
{\mathcal L}_{{\mathcal U}^{spin-0}}=  \frac{\lambda_{ij}^{S}}{\Lambda_{\mathcal U}^{d_{\mathcal U}-1}} \bar{f}_i f_j {\mathcal O}_{\mathcal U} +\frac{\lambda_{ij}^{P}}{\Lambda_{\mathcal U}^{d_{\mathcal U}-1}} \bar{f}_i \gamma^5 f_j {\mathcal O}_{\mathcal U},
\end{equation}
where $\lambda_{ij}^{{S,P}}=C_{{\mathcal O}_{\mathcal U}} \Lambda_{\mathcal U}^{d_{{\mathcal B}{\mathcal Z}}}/{\mathcal M}_{\mathcal U}^{d_{SM}+d_{{\mathcal B}{\mathcal Z}}-4}$ stands for the respective coupling constant.
Constraints on the coupling constant associated with the $\tau$ lepton have been obtained from the LFV decay $\tau \to 3 \mu$ \cite{Moyotl:2011yv} and the muon anomalous magnetic moment \cite{Hektor:2008xu}. As far as the unparticle propagators are concerned, they are constructed using scale invariance and the spectral decomposition formula. The propagator for a spin-0 unparticle can be written as

 \begin{equation}
\Delta_F(p^2)= \frac{A_{d_\mathcal U}}{2\sin(d_\mathcal U \pi)} (-p^2-i\epsilon)^ {d_{\mathcal U}-2}, \qquad \text{where} \qquad A_{d_\mathcal U}=\frac{16\pi^2\sqrt{\pi}}{(2\pi)^{2d_\mathcal U}}\frac{\Gamma(d_\mathcal U+\frac{1}{2})}{\Gamma(d_\mathcal U-1)\Gamma(2d_\mathcal U)}. \label{unpro}
\end{equation}
normalize the spectral density \cite{Cheung:2007ap}.

\section{The electric and weak electric dipole moment of fermions}

The effective Lagrangian  describing  the EDM and the WEDM is ${\mathcal L}^{\text{spin}-1/2} =-\frac{i}{2} \bar{f} \sigma_{\mu\nu} \gamma_5 f(d_f F^{\mu\nu}_\gamma+d_f^{W} F^{\mu\nu}_Z)$, where $F^{\mu\nu}_\gamma$ and $F^{\mu\nu}_Z$ are the electromagnetic and weak stress tensors, respectively.
The EDM and WEDM arise at the loop level and can be extracted from the matrix element $ie\bar{\mathrm{u}}(p') \Gamma^\mu_V \mathrm{u}(p)$, where $ \Gamma_V^\mu$ is given by:

\begin{eqnarray}
\Gamma^\mu_V(q^2)= F_A(q^2)(\gamma^\mu \gamma_5 q^2-2{m_f} \gamma_5 q^\mu)+F_1(q^2) \gamma^\mu +F_2(q^2)i\sigma^{\mu\nu} q_\nu+F_3(q^2)\sigma^{\mu\nu} \gamma_5 q_\nu, \label{vad}
\end{eqnarray}
with $q=p'-p$ the four-momentum of the neutral gauge boson $V=Z$, $\gamma$. Then,  the EDM and WEDM of a fermion are given by  $d_f=-eF_3(q^2=0)$ and $d_f^W=-eF_3(q^2=m_Z^2)$, respectively. The flavor changing interaction given by Eq. (\ref{lup}) have been considered to  calculate the one-loop $\bar{f}fV$ vertex via Feynman parameters. The result for the WEDM can be written as

\begin{eqnarray}
d_f^{{W}}(d_\mathcal U) =\frac{ -  e g_V^f A_{d_\mathcal U}}{16 \pi^2m_f\sin{(d_\mathcal U} \pi)} \sum_{i=e,\mu,\tau} \text{Im}\big( {\lambda_{f i}^P}^* \lambda_{f i}^S \big) \sqrt{r_i} \bigg( \frac{m_i^2}{\Lambda_{\mathcal U}^2} \bigg)^{d_\mathcal U-1}  \int_0^1dx  \int_0^{1-x}dy (1-x) H(d_\mathcal U,r_{i},x_Z,x,y),\label{dw1}
\end{eqnarray}
where  $r_{i}=m_f^2/m_i^2$ and $g_V^f$ is the vector coupling constants of the $Z$ gauge boson to pair of fermions. The dimensionless function $H(d_\mathcal U,r_{i},x_Z,,x,y)=x^{1-d_\mathcal U}\left(r_{i}x_Z (x+y-1)y+(1-x)(1-r_{i}x)\right)^{d_\mathcal U-2}$ was introduced, with $x_Z=m_Z^2/m_f^2$ Furthermore, $\text{Im}\big( {\lambda_{fi}^P}^* \lambda_{fi}^S \big)=|\lambda_{fi}^S||\lambda_{fi}^P| \sin(\theta_{fi}^S-\theta_{fi}^P)$. The fermion WEDM only receives contributions from the vector coupling $g_V^f$ and it is necessary that $f\ne i$, which is expected as this property violates CP.  As a cross-check for our calculation, from Eqs.  (\ref{dw1}) the fermion EDM reported in Ref. \cite{Moyotl:2011yv} can be obtained after the replacements $x_Z=0$, $g_A^f=0$ and $g_V^f=Q_f$ are done. Here $Q_f$ is the fermion electric charge in units of $e$.

\section{Numerical analysis and discussion}

The analysis of mono-photon production plus missing transverse energy, $e^+e^-\to\gamma +X$, at the LEP was used in Ref. \cite{Cheung:2007ap} to impose a bound on the scale $\Lambda_{\mathcal U}$ as a function of $d_{\mathcal U}$. They considered the 95 \%C. L. limit  $\sigma(e^+e^-\to\gamma +X)\simeq 0{.}2$ pb obtained at $\sqrt{s}=207$ GeV by the L3 Collaboration. It was found that this limit requires $\Lambda_{\mathcal U}\ge 660$ TeV for $d_\mathcal U=1{.}4$ and $\Lambda_{\mathcal U}\ge 1{.}35$ TeV for  $d_\mathcal U= 2$. Stronger limits were obtained by the CMS collaboration using the data for mono-jet production plus missing transverse energy at the LHC for $\sqrt{s}=7$ TeV and an integrated luminosity of 35 pb$^{-1}$. Such data require  $\Lambda_{\mathcal U}\ge 10$ TeV for $d_\mathcal U=1{.}4$ and $\Lambda_{\mathcal U}\ge 1$ TeV for  $d_\mathcal U=1{.}7$ \cite{Chatrchyan:2011nd}. In summary, the region $d_\mathcal U< 1{.}4$ is strongly constrained as very large values of $\Lambda_{\mathcal U}$ are required. Depending on the relative CP-violating phase $\theta_{\tau\mu}^S-\theta_{\tau\mu}^P$, the $\tau$ EDM or WEDM can be negative or positive, which poses no problem as the experimental bound also comprehends negative values. In order to analyze the unparticle contribution to the $\tau$ EDM and WEDM, no specific values for $\text{Im}\big( {\lambda_{\tau\mu}^P}^* \lambda_{\tau\mu}^S \big)$ will be considered.

\begin{figure}[!hbt]
\centering
\includegraphics[width=8.3cm]{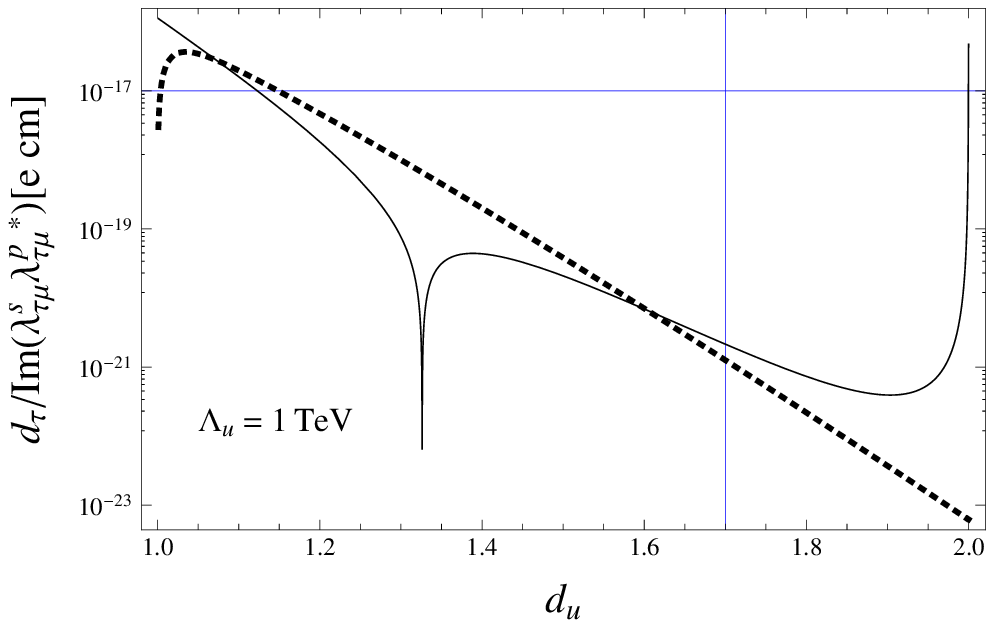}
\includegraphics[width=8.3cm]{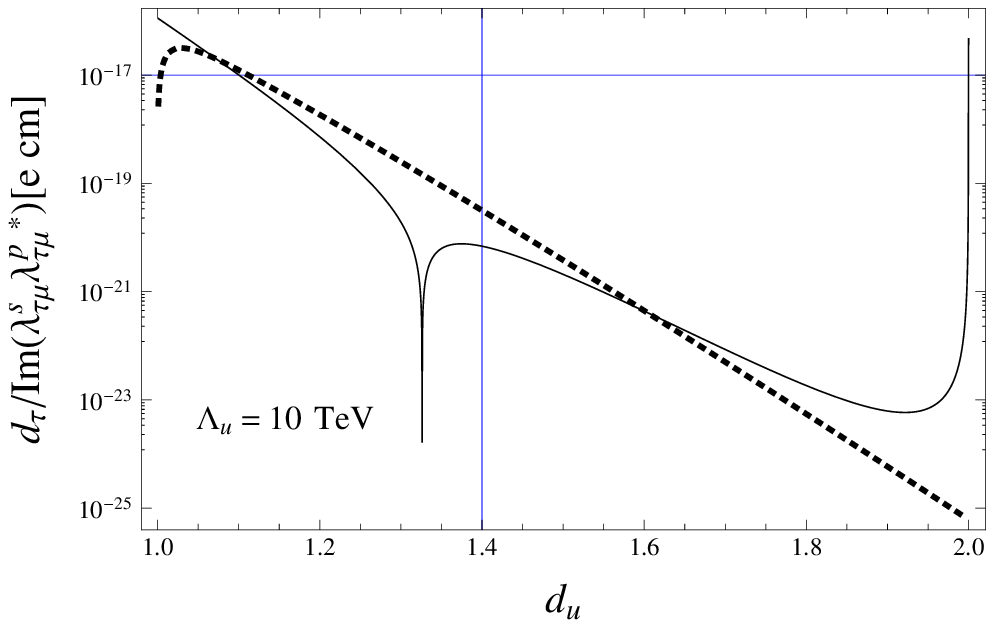}
\caption{Absolute values of the real (solid line) and imaginary (dotted line) parts of the contribution from a spin-0 unparticle to the $\tau$ EDM as a function of the scale dimension $d_\mathcal U$ for two values of $\Lambda_\mathcal U$. The horizontal line is the SM contribution, and the vertical line represents the lower bound obtained by the CMS collaboration \cite{Chatrchyan:2011nd}.}
\label{dem1-10}
\end{figure}

{\bf Tau electric dipole moment.} In Fig.  \ref{dem1-10} I have plotted the absolute values of Re($d_\tau$) and Im($d_\tau$) from a spin-0 unparticle as a function of the scale $d_\mathcal U$ for $\Lambda_{\mathcal U}=1$ TeV  and $\Lambda_{\mathcal U}=10$ TeV. A detailed analysis allows one to conclude that  there is a change in the sign of Re($d_\tau$) at $d_\mathcal U\simeq 1{.}325$, whereas that Im($d_\tau$) is always positive. In the allowed region, both the real and imaginary parts are positive, although the CP-violating phase can give an additional change of sign. It is also interesting that Re($d_\tau$) diverges when $d_\mathcal U \to 2$, but Im($d_\tau$) is negligibly small. Therefore, around  $d_\mathcal U=2$, the $\tau$ EDM is almost real and also reaches its largest size. In general, the spin-0 unparticle contribution to $d_\tau$ can be above the SM prediction \cite{Hoogeveen:1990cb} as long as $\text{Im}\big( {\lambda_{\tau\mu}^P}^* \lambda_{\tau\mu}^S \big)$ is not too small. As far as other SM extensions are concerned, in an extension of the minimal supersymmetric standard model with vectorlike multiplets, the contributions to the $\tau$ EDM arise at the one loop level from loops carrying  $W$ gauge bosos, charginos (${\tilde{\chi}}_i^\pm$) or neutralinos (${\tilde{\chi}}_i^0$). Since these particles are heavier than the $\tau$ lepton, their contributions to the EDM are purely real and  have values ranging from $d_\tau\simeq 6{.}5\times10^{-18}$ e cm to $d_\tau\simeq 3{.}0\times10^{-23} $ e cm \cite{Ibrahim:2010va}. In contrast, the unparticle contribution $d_\tau^{\mathcal U}$ is almost real at $d_\mathcal U\simeq 2$, where it can reach values of the order of  $10^{-18}$ e cm, though it tends to be smaller for other $d_\mathcal U$ values.  The $\tau$ EDM has also been studied in other SM extensions, but the respective predictions  were found to be very small. This is the case of the framework of the Fritzsch-Xing lepton mass matrix, in which $|d_\tau|<2{.}2\times 10^{-25}$ e cm \cite{Fritzsch:1999ee,Huang:1999vb}.

\begin{figure}[!hbt]
\centering
\includegraphics[width=8.3cm]{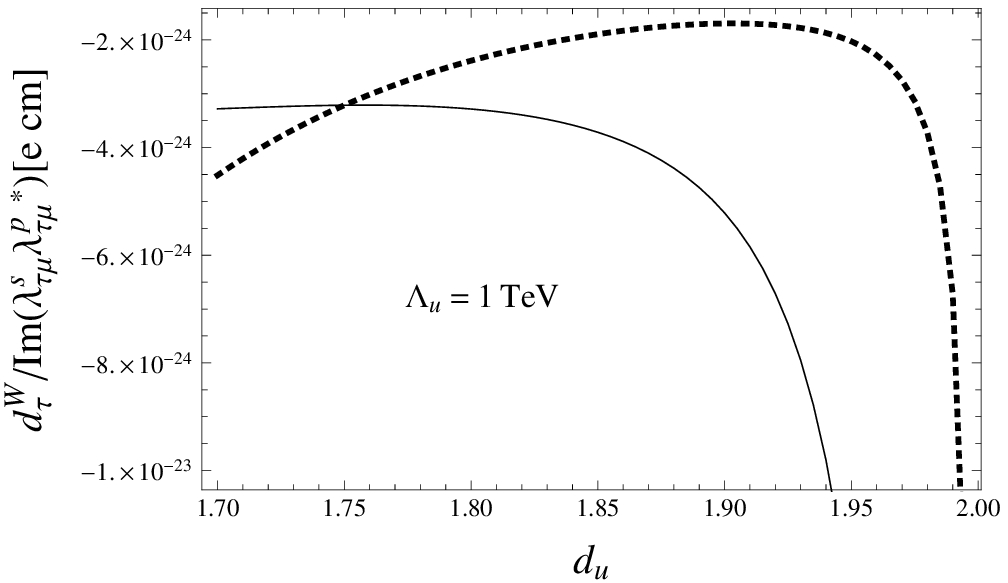}
\includegraphics[width=8.3cm]{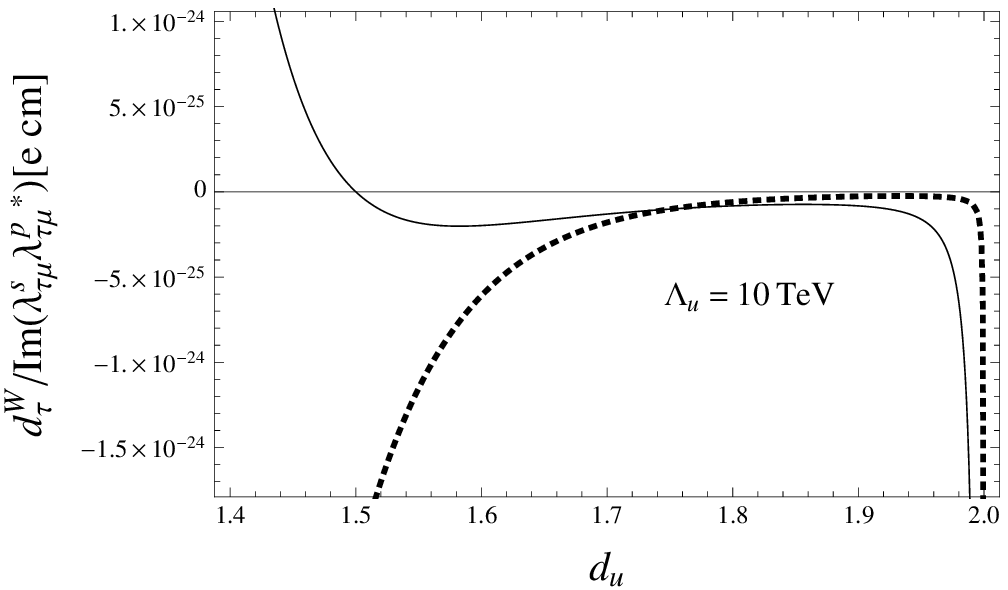}
\caption{The same as in Fig \ref{dem1-10}, but for the $\tau$ WEDM.}
\label{dwt}
\end{figure}

{\bf Tau weak electric dipole moment.} The analysis of  $H(d_\mathcal U,r_{i},x_Z,,x,y)$ suggests  that the $\tau$ WEDM is expected to be smaller than the EDM due to the term proportional to $x_Z$. I show in Fig.  \ref{dwt} the behavior of Re($d_\tau^W$) and Im($d_\tau^W$) induced by a spin-0 unparticle as a function of $d_\mathcal U$, for two values of $\Lambda_\mathcal U$.  As was anticipated, this property shows a behavior similar to that of the $\tau$ EDM, though it has a smaller magnitude and opposite sign. When $\Lambda_{\mathcal U}=1$ TeV, both the real and imaginary contributions are negative, but when $\Lambda_{\mathcal U}=10$ TeV, the imaginary part is negative whereas the real part changes from positive to negative at $d_\mathcal U \simeq 1{.}5$, where the $\tau$ WEDM is almost imaginary, i.e. $d_\tau^{W}\simeq-i\text{Im}\big( {\lambda_{\tau\mu}^P}^* \lambda_{\tau\mu}^S \big)2{.}2\times10^{-24}$ e cm. Contrary to behavior of the EDM, both  the real and imaginary parts of the WEDM diverge when $d_\mathcal U\to 2$.

\section{Conclusions}

The EDM and WEDM has been studied in the framework of unparticle physics. In this work I only considered the contribution from a spin-0 unparticle and for the respective coupling constants I used the most recent CMS bounds from mono-jet production plus missing transverse energy at the LHC. In the most promising scenario, the unparticle contribution to the EDM can be larger than the contributions predicted by the SM and an extension of the minimal supersymmetric standard model with vectorlike multiplets. As far the $\tau$ WEDM  is concerned, the contributions from a spin-0 unparticle are smaller than the respective contributions to its electromagnetic analogue, although they are larger than the SM contributions, tough much smaller than the current experimental limits \cite{PhysRevD.86.013014}.

\begin{acknowledgments}

 I would like to thank IF-BUAP for partial support to cover my attendance to this conference.  I am grateful to the conference organizers for their kind invitation to present my research work and their kind hospitality. I also would like to thank Lizeth R. B. for reading this manuscript and making useful suggestions.

\end{acknowledgments}

\bibliography{FPCP2012-64Moyotlv1}

\end{document}